\documentclass[oldversion]{aa}
\usepackage{amsmath}
\usepackage{amssymb}
\usepackage{natbib}
\usepackage{graphicx}
\usepackage{txfonts}
\usepackage[english]{babel}

\begin{document}

\title{The shape and composition of interstellar silicate grains}

\author{M. Min\inst{1}
\and L.~B.~F.~M. Waters\inst{1,2}
\and A. de Koter\inst{1}
\and J.~W. Hovenier\inst{1}
\and L.~P. Keller\inst{3}
\and F. Markwick-Kemper\inst{4}
}

\offprints{M. Min, \email{mmin@science.uva.nl}}

\institute{Astronomical institute Anton Pannekoek, University of Amsterdam, Kruislaan 403, 
1098 SJ  Amsterdam, The Netherlands
\and Instituut voor Sterrenkunde, Katholieke Universiteit Leuven, Celestijnenlaan 200B, 
3001 Heverlee, Belgium
\and Mail Code KR, NASA Johnson Space Center, Houston, TX 77058
\and University of Virginia, Department of Astronomy, PO Box 400325, Charlottesville VA 22904-4325
}

\date{Accepted for publication in A\&A, November 6, 2006}

\abstract
{We investigate the composition and shape distribution of silicate dust grains in the  interstellar medium. The effects of the amount of  magnesium and iron in the silicate lattice are studied in detail. We fit the spectral shape of the  interstellar 10$\,\mu$m extinction feature as observed towards the galactic center using various particle shapes and dust materials.  We use  very irregularly shaped coated and non-coated porous Gaussian Random Field particles as well as a statistical  approach to model shape effects. For the dust materials we use amorphous and crystalline  silicates with various composition as well as silicon carbide (SiC). The results of our  analysis of the 10$\,\mu$m feature are used to compute the shape of the 20$\,\mu$m  silicate feature and to compare this with observations of this feature towards the  galactic center. By using realistic particle shapes to fit the interstellar extinction spectrum we are, for  the first time, able to derive the magnesium fraction in interstellar silicates. We find  that the interstellar silicates are highly magnesium rich  ($\mathrm{Mg/(Fe+Mg)}>0.9$) and that the stoichiometry lies between pyroxene and olivine type silicates ($\mathrm{O/Si}\approx3.5$). This composition is not consistent with that of the glassy material found in GEMS in interplanetary dust particles indicating that the amorphous silicates found in the Solar system are, in general, not unprocessed remnants from the interstellar medium. Also, we find  that a significant fraction of silicon carbide ($\sim$3\%) is present in the interstellar  dust grains. We discuss the implications of our results for the formation and evolutionary  history of cometary and circumstellar dust. We argue that the fact that crystalline silicates in cometary and circumstellar grains are almost purely magnesium silicates is a natural consequence of our findings that the amorphous silicates from which they were formed were already magnesium rich. }


\maketitle

\section{Introduction}
\label{sec:introduction}

Understanding the composition of silicate dust in the Interstellar Medium (ISM) is crucial  for studies of circumstellar and cometary dust and for the processing of dust in interstellar space. For example, in order to understand the  way in which dust is modified in protoplanetary disks during the process of disk  dissipation and planet formation, it is important to know the original composition of the  material that entered the disk. It is reasonable to assume that this material had a  composition similar to that found in the diffuse interstellar medium. The amount of iron and magnesium in the silicates contains valuable information on  the origin of the grains and is an important probe of the processing history. When the silicates are crystalline, i.e. they have a long range order in the lattice, the magnesium content can be readily obtained from the spectral  positions of sharp solid state emission features. In this way it has been shown that  cometary and circumstellar crystalline silicates are highly magnesium rich, i.e.  $\mathrm{Mg/(Fe+Mg)}>0.9$ \citep[see e.g.][]{1998A&A...339..904J, 2001A&A...378..228F, 2003A&A...399.1101K}. For amorphous silicates, i.e. silicates with a disordered lattice structure, the spectral features are less pronounced which makes it harder to determine the composition from spectroscopy. In particular, it is hard to disentangle effects of grain shape and  composition on the spectral profile. In previous studies the interstellar silicates have been modeled using amorphous silicates  containing approximately equal amounts of iron and magnesium \citep[see e.g.][]{1984ApJ...285...89D, 2001ApJ...550L.213L, 2004ApJ...609..826K}. By including amorphous silicates with slightly different compositions \citet{2004ApJ...609..826K} concluded that the composition is indeed most likely $\mathrm{Mg/(Fe+Mg)}\approx0.5-0.6$. If the grain properties derived in these studies are correct it is difficult to understand the high  magnesium content of crystalline silicates in protoplanetary disks and solar system comets. The reason is that the majority of crystalline silicates is expected to be formed by thermal annealing of amorphous silicates, preserving the composition of the amorphous silicates, while only a small fraction is expected to originate from gas-phase condensation, producing magnesium rich crystals \citep{2004A&A...413..571G}.

\begin{figure*}[!t]
\resizebox{\hsize}{!}{\includegraphics[height=10cm]{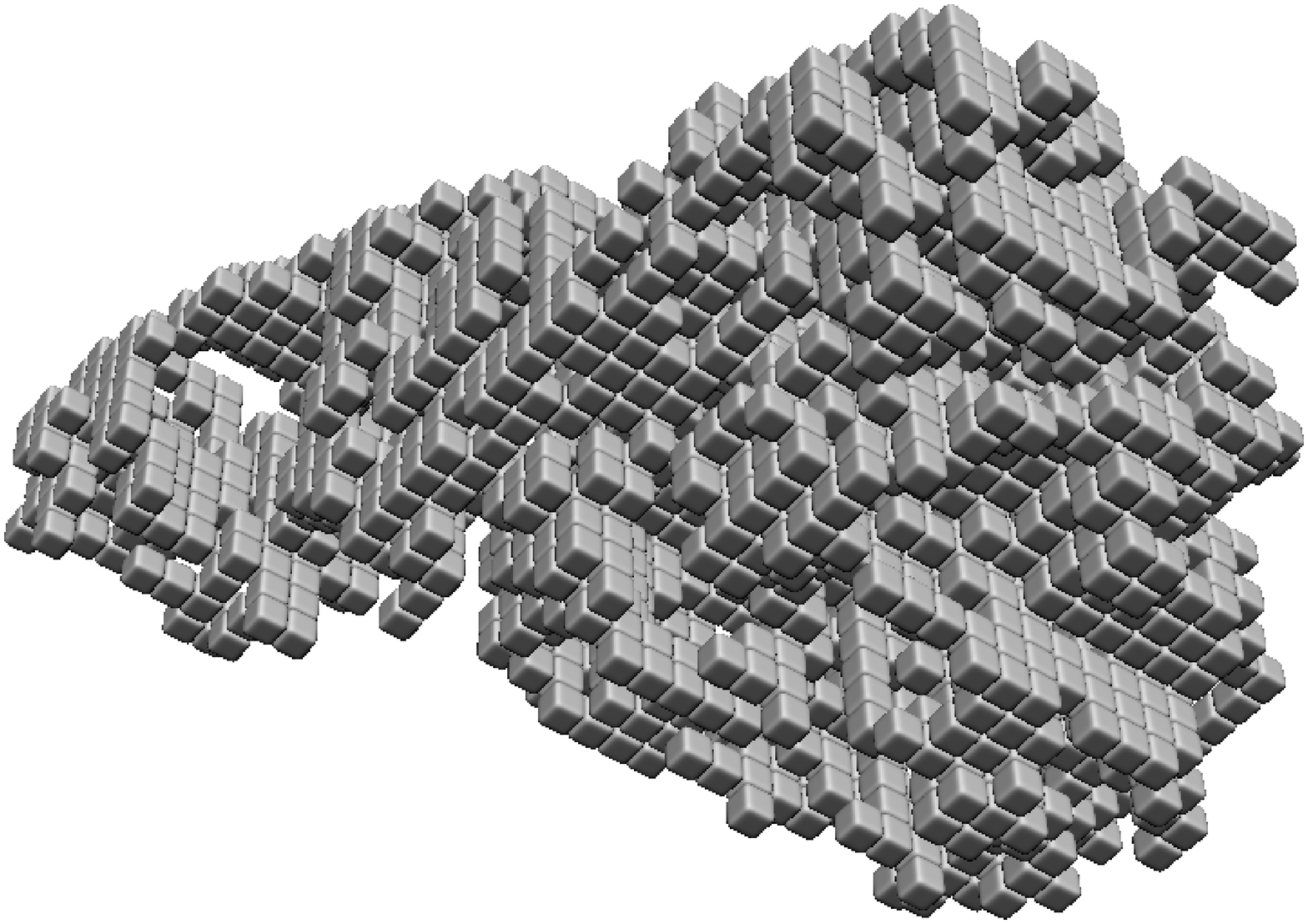}\includegraphics[height=10cm]{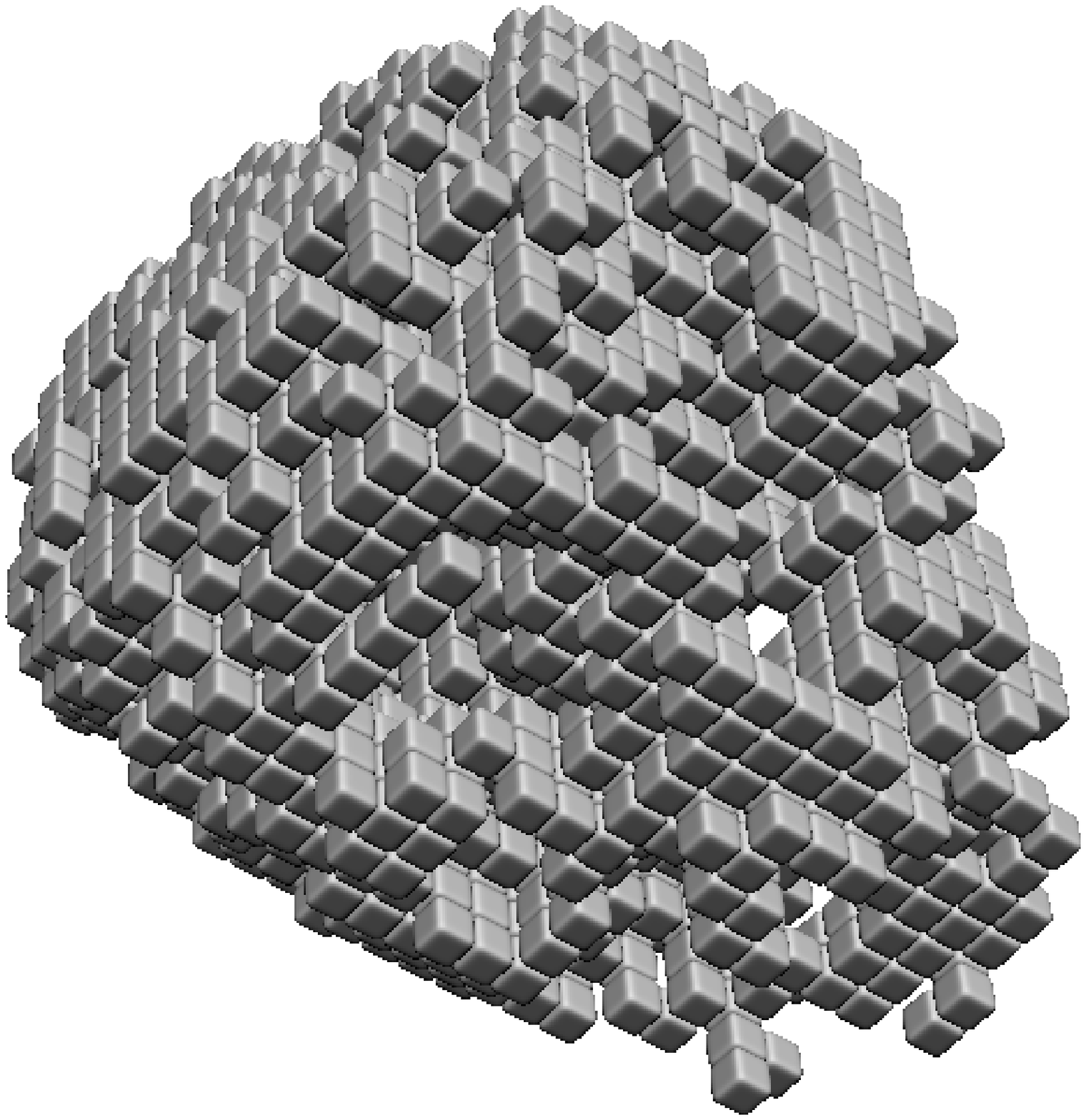}\includegraphics[height=10cm]{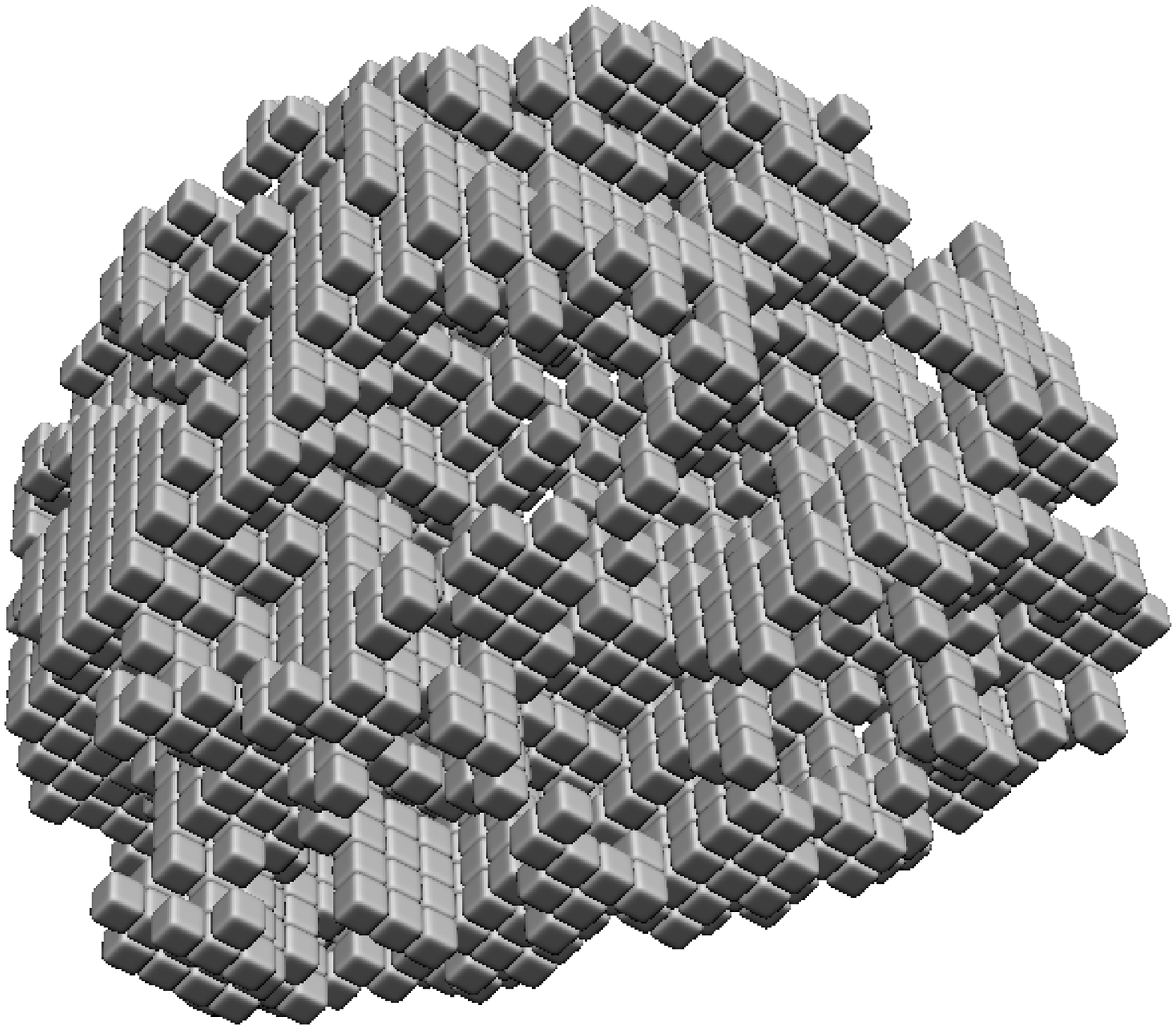}}
\caption{Typical examples of porous Gaussian Random Field particles generated using the method described in the text. Every small cube seen in this figure represents a volume element according to the discretization of space used for the construction of the Gaussian random field. The grains shown here are typical examples from the ensemble of 38 porous GRF particles used to compute the extinction spectra.}
\label{fig:GRF particles}
\end{figure*}

In the above mentioned studies the 10$\,\mu$m ISM extinction feature has been modeled assuming that  the optical properties of the grains can be represented by those of perfect homogeneous  spheres. This allows for fast and simple computations. The fit to the 10$\,\mu$m ISM extinction  feature using homogeneous spheres is surprisingly good, and it is therefore tempting to  assume that the derived dust properties are correct. There are, however, several  indications that the assumption of homogeneous spheres cannot be adequate.
First there is the interstellar polarization. A homogeneous sphere of optically inactive  material, like silicates, cannot produce linear polarization through extinction. Since the  10$\,\mu$m silicate feature is clearly visible in the spectral dependence of the degree of  interstellar linear polarization \citep{1985ApJ...290..211L, 1986MNRAS.218..363A, 1989MNRAS.236..919A}, the silicates have to contribute to it. The most natural explanation  for this is extinction by aligned, nonspherical silicate grains. The interstellar polarization alone, however, does not preclude using the spherical grain model for explaining the extinction spectrum. One might assume that spherical grains can be used to model the average extinction properties of slightly elongated particles. In addition, it appears that only very modest deviations from a perfect sphere are needed to produce the required polarization \citep[an aspect ratio of $\sim2$ seems sufficient, see e.g.][]{1985ApJ...290..211L, 1993A&A...280..609H}.
However, it has been shown recently that the 20$\,\mu$m silicate extinction band cannot be  accurately reproduced by using homogeneous spherical dust grains  \citep{2006ApJ...637..774C}.
In addition it is clear from several studies that effects of particle shape on the extinction spectrum of small particles can be very large.
Therefore, a detailed re-analysis of the 10$\,\mu$m ISM extinction feature using more  realistic grain models is needed.

In this paper we study the effects of grain non-sphericity on the derived composition and  lattice structure of interstellar grains. We do this by computing the optical properties  of an ensemble of so-called porous Gaussian Random Field (GRF) particles and by applying a  statistical approach to account for particle shape effects. For this study the exact  choice of the shapes of the nonspherical particles is not crucial since it is shown by  \citet{2003A&A...404...35M} that the optical properties of various classes of nonspherical  particles are quite similar, and it is mainly the perfect symmetry of a homogeneous sphere  that causes a different spectral signature.
From our analysis of the 10$\,\mu$m spectral extinction feature we can compute the  expected shape of the 20$\,\mu$m feature. We compare this to the observed extinction  spectrum obtained by \citet{2006ApJ...637..774C}.

The structure of the paper is as follows. In section~\ref{sec:Extinction spectra} we  explain in some detail the effects of particle shape and composition on the shape of the  10$\,\mu$m spectral profile. The procedure to fit the 10$\,\mu$m feature and the results  of this fit are presented in section~\ref{sec:ISM grains}. Implications of our results are  discussed in section~\ref{sec:Discussion}. Finally, in section~\ref{sec:Conclusions} we  summarize the conclusions of the paper.

\section{Extinction spectra}
\label{sec:Extinction spectra}

\subsection{Effects of grain shape}

The shape of a dust grain has a large influence on the shape of the absorption and  extinction spectrum it causes. This effect has been described by e.g. \citet{BohrenHuffman, 2001A&A...378..228F, 2003A&A...404...35M, 2006A&A...445..167V}. In many cases one needs  to consider irregularly shaped dust grains in order to reproduce the infrared spectra  observed from circumstellar, cometary or laboratory particles correctly \citep[see  e.g.][]{2002A&A...390..533H, 2003A&A...401..577B, 2003ApJ...595..522M,  2003A&A...404...35M}. It can be shown that the absorption spectrum caused by homogeneous spherical particles is  very different from that caused by other particle shapes. This  difference is much larger than the differences due to various nonspherical particles  shapes \citep{2003A&A...404...35M}. The perfect symmetry of a homogeneous sphere causes a resonance with a spectral shape and position very different from resonances observed in the spectra of all other particle shapes. This emphasizes why it is puzzling that the interstellar 10$\,\mu$m extinction feature can be fitted almost perfectly using homogeneous amorphous silicate spheres. In this paper we  will consider three types of particle shapes, homogeneous spheres, porous Gaussian Random  Field (GRF) particles and a statistical ensemble of simple particle shapes to represent  irregularly shaped particles.

\subsubsection{Gaussian Random Field particles}

\begin{figure}[!t]
\resizebox{\hsize}{!}{\includegraphics[height=10cm]{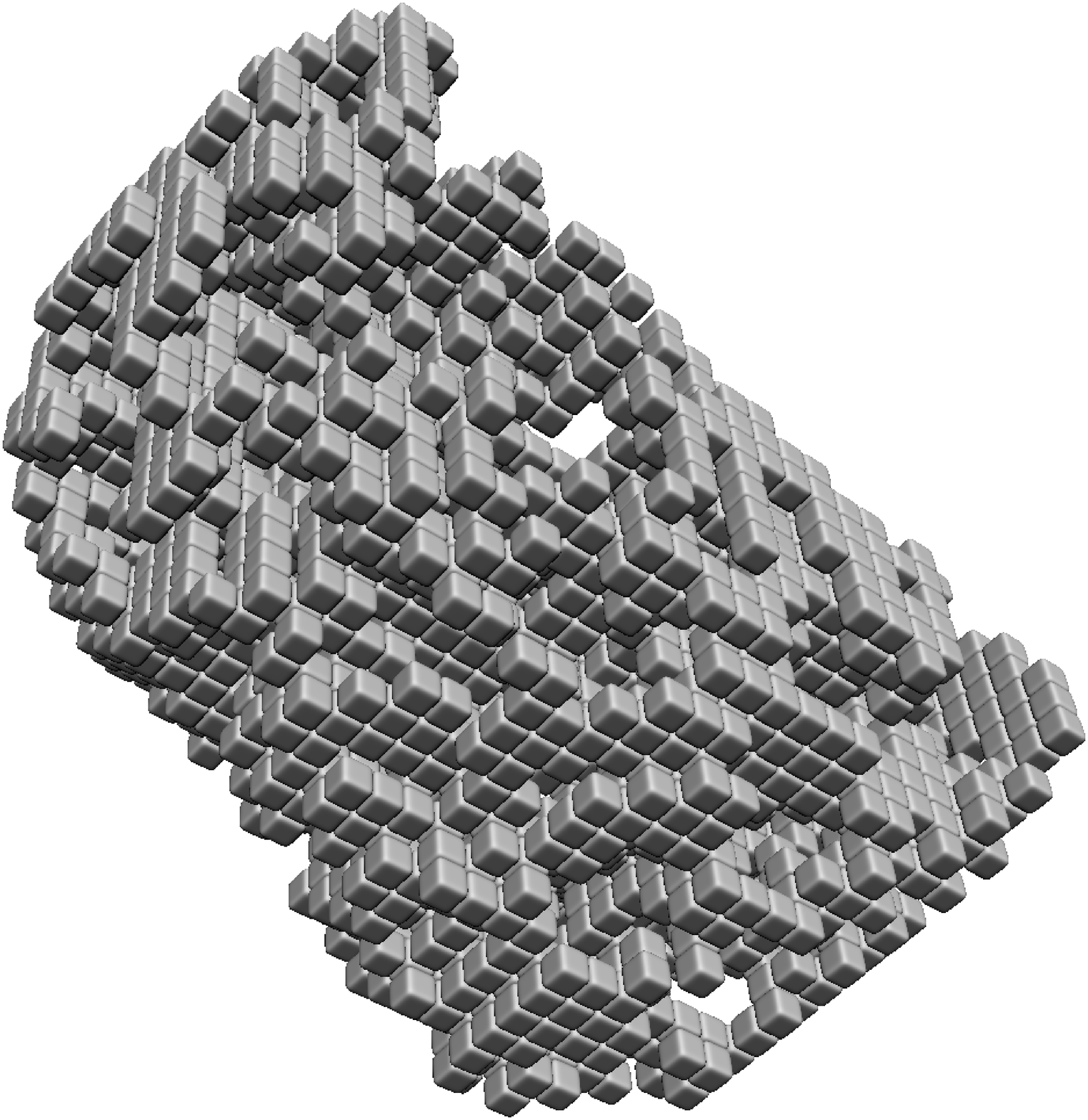}\includegraphics[height=10cm]{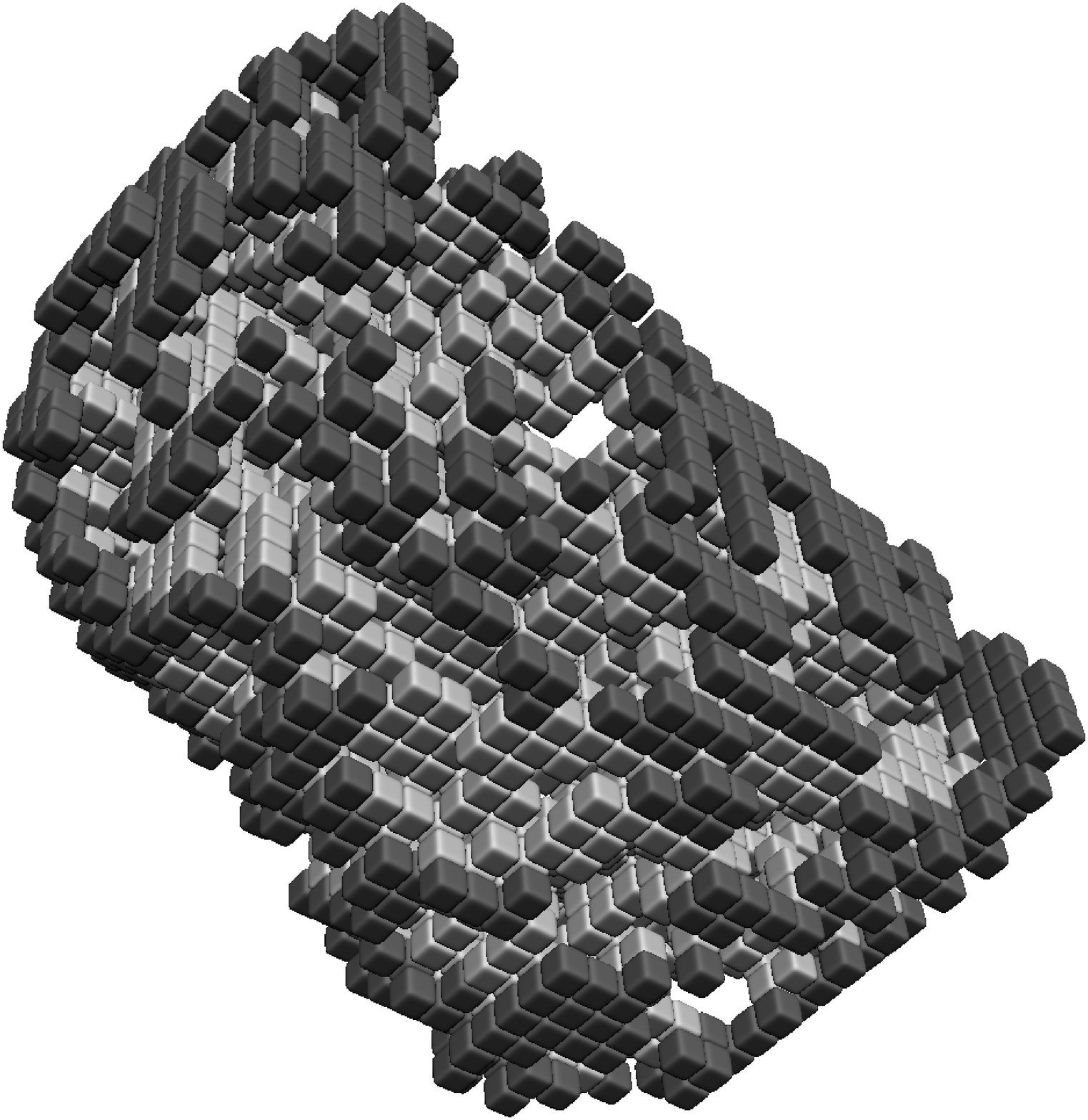}}
\caption{A porous GRF particle without (left) and with (right) a carbon mantle. Both particles are created from the same Gaussian Random Field. The dark grey cubes in the right picture represent the amorphous carbon mantle which takes up 30\% of the material of the grain.}
\label{fig:CoatedGRF}
\end{figure}

Several attempts have been made to properly model irregularly shaped particles. One of the  most widely used classes of irregularly shaped particles is perhaps that of Gaussian  Random Spheres \citep{1996JQSRT..55..577M}. These particle shapes have the advantage that  they are easily generated numerically. However, the resulting particles do not appear to  resemble accurately realistic particle shapes thought to be representative for those in  the ISM like, for example, Interplanetary Dust Particles \citep[IDPs; see  e.g.][]{DustCatalog}. A more promising class of particles in this respect might be the  Gaussian Random Field (GRF) particles \citep[see e.g.][]{2003JQSRT..78..319G,  2005Icar..173...16S}. This class of particles covers a wide range of very irregular shapes. In addition, the particles are also relatively easily generated numerically.

In short, a GRF particle is constructed as follows. First, space is divided into small  volume elements. Every volume element is then given a value which is correlated with the  values of surrounding volume elements according to a lognormal distribution. The resulting  three dimensional field is called a GRF. A threshold level is introduced (in our case 0.5  times the maximum value). All volume elements with a value of the GRF above the threshold  are inside the particle, while all other volume elements are considered vacuum. In this  way an ensemble of particles is created. Since we considered the resulting particles to be  still rather smooth, we developed a way to make the particles more irregular. This was done by creating a second GRF with a much smaller correlation length. Where this second field has  a value below a certain threshold, we place vacuum voids. In this  way, irregularly shaped voids are created inside the particles, making them porous, and on the edge of the particles, making them more rough. We will refer to these particles as porous GRF particles. Some pictures of typical examples of porous GRF particles are shown in Fig.~\ref{fig:GRF particles}. Every small cube seen in this figure is a volume element according to the discretization of space chosen for the construction of the GRF.  We construct an ensemble of 38 porous GRF particles and average the extinction spectra over this ensemble. For details on the algorithm to construct porous Gaussian Random Field particles, see Appendix~\ref{app:GRF}.

In order to study the effect of mixed carbon/silicate grains we also consider coated porous GRF particles. These are created in such a way that a fixed abundance of 30\% of the outer layer of the particle consists of amorphous carbon. This can be done easily using porous GRF particles by taking two thresholds in the GRF from which the particle is created. For points in the grain where the value of the GRF is $>0.5$ and smaller than the second threshold, we have the carbon mantle. For points in the grain where the GRF is larger than the second threshold, we have the core of the grain. The value of the second threshold is chosen such that the mantle contains 30\% of the total grain material volume. A picture of a coated porous GRF particle is shown in Fig.~\ref{fig:CoatedGRF}.

\subsubsection{Distribution of Hollow Spheres}

\begin{figure}[!t]
\resizebox{\hsize}{!}{\includegraphics{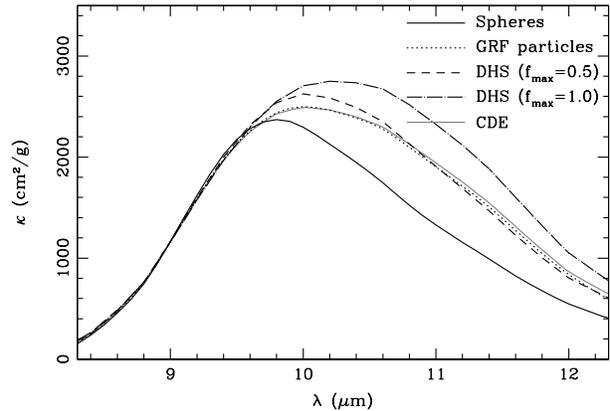}}
\caption{The mass extinction coefficient as a function of wavelength in the $10\,\mu$m 
region for amorphous silicate particles with an olivine type stoichiometry and $x=\mathrm{Mg/(Fe+Mg)}=0.5$ 
(i.e. MgFeSiO$_4$) with various shape distributions. The feature as obtained using the widely employed Continuous Distribution of Ellipsoids (CDE; grey line) is shown for comparison.}
\label{fig:Shape}
\end{figure}

When modeling the optical properties of irregularly shaped particles in astrophysical  environments we have to keep in mind that we always observe the average optical properties  of an ensemble of particles with various shapes. Therefore, we are not interested in the  specific characteristic optical properties of a single particle, but rather in the average  optical properties of the entire ensemble. Once we realize this, it is a logical step to  use a \emph{statistical approach} to compute the average optical properties of an ensemble  of irregularly shaped particles. In this statistical approach we assume that the optical  properties of an ensemble of irregularly shaped particles can be modeled in a statistical  way by using the average optical properties of a distribution of simple particle shapes the individual particles of  which need not necessarily be realistically shaped. For particles  in the Rayleigh domain, i.e. particles much smaller than the wavelength both inside and  outside the particle, it was shown analytically that for all ensembles of particle shapes  an ensemble of spheroidal shapes exists with identical absorption and extinction cross  sections \citep{2006JQSRT..97..161M}. One advantage of using the statistical approach is  that the computation time is limited since it avoids computations of large ensembles of  complex shaped particles. Another advantage is that by using simple particle shapes, the  number of parameters used to describe the particles is limited. This helps determining the  main characteristics of the particles and avoids over-interpretation.

A particularly successful implementation of the statistical approach is given by the  Distribution of Hollow Spheres \citep[DHS;][]{2003A&A...404...35M, 2005A&A...432..909M}. With the DHS it is possible to reproduce measured and observed spectra of small particles while computations can be performed easily and fast for arbitrary particle sizes \citep[for an application of this method see e.g.][]{2005Icar..179..158M}. The DHS considers hollow spherical particles, averaging the optical properties uniformly  over the fraction of the total volume occupied by the central vacuum inclusion, $f$, over  the range $0<f<f_\mathrm{max}$ while keeping the material volume of the particles  constant. The value of $f_\mathrm{max}$ reflects the degree of irregularity of the  particles considered. We would again like to stress here that by taking this approach we do \emph{not} consider the real particles to be hollow spherical shells. Rather, it is assumed that the optical properties of an ensemble of realistically shaped, irregular particles can be represented by the average properties of an ensemble of hollow spheres in a statistical sense.

\subsubsection{The amorphous silicate feature}

The interstellar silicates  causing the 10$\,\mu$m silicate extinction feature are mostly amorphous. \citet{2004ApJ...609..826K, 2005ApJ...633..534K} placed an upper limit on the amount of crystalline silicates of 2.2\%. In addition, the  particles in the interstellar medium are very small ($\lesssim 0.3\,\mu$m). At  $\lambda=10\,\mu$m this implies that the particles are much smaller than the wavelength  both inside and outside the particle, i.e. they are in the Rayleigh domain. For particles  in the Rayleigh domain the actual particle size is not important for the shape of the  absorption and extinction spectra. In order to compute the spectra of porous GRF particles  we employed the method of \citet{2006JQSRT..97..161M} which allows for rapid computation  of absorption and extinction spectra of arbitrarily shaped particles in the Rayleigh  domain. For the spherical particles we use Mie theory \citep{Mie}. To compute the optical  properties of the distribution of hollow spheres we use a simple extension of Mie theory  \citep{1981ApOpt..20.3657T}.

Fig.~\ref{fig:Shape} shows the mass extinction coefficient, $\kappa$, as a function of  wavelength for amorphous silicate grains with an olivine type stoichiometry and an equal  amount of magnesium and iron (see also section~\ref{sec:composition}) for homogeneous  spherical particles, for an ensemble of porous GRF particles, for two distributions of  hollow spheres with $f_\mathrm{max}=0.5$ and $1.0$, and for a Continuous Distribution of Ellipsoids (CDE). It is clear that the shape and position of the feature has a strong dependence on grain shape. In general the spectral extinction features caused by  irregularly shaped particles are much broader and shifted towards the red with respect to  those caused by homogeneous spherical particles \citep[see also][]{2003A&A...404...35M}. This effect becomes stronger when deviating more from the perfect homogeneous sphere, as can be seen when increasing $f_\mathrm{max}$.

\subsection{Effects of grain composition}
\label{sec:composition}

\begin{figure}[!t]
\resizebox{\hsize}{!}{\includegraphics{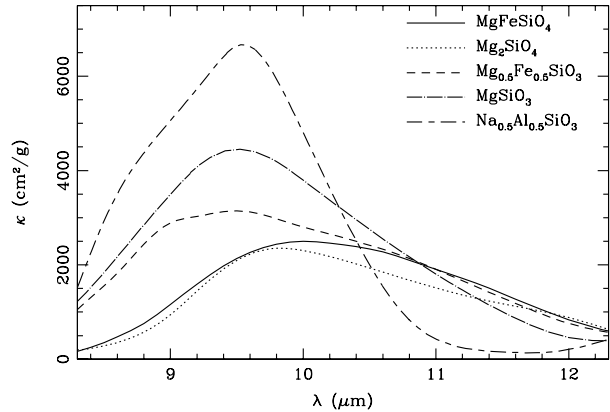}}
\caption{The mass extinction coefficient as a function of wavelength in the $10\,\mu$m 
region for porous GRF particles of various compositions.}
\label{fig:Composition}
\end{figure}

Although silicates with various compositions all display a spectral feature in the  $10\,\mu$m region due to the Si-O stretching mode, there are differences in the spectral  appearance that allow to constrain the composition of the silicates. In our analysis we consider amorphous silicates with various compositions, the crystalline  silicates forsterite and enstatite, and silicon carbide (SiC). Although we expect that carbonaceous species and metallic iron are important constituents of interstellar dust, we have no constraints on their abundances from spectroscopy around $10\,\mu$m since in this wavelength range they have a smooth, featureless extinction spectrum. Therefore, they are not detectable in our analysis. The various components we use will be discussed below.

Silicates are composed of linked SiO$_4$ tetrahedra with cations inbetween (e.g. Mg, Fe,  Na, Al). A silicate can have various stoichiometries depending on the number of oxygen  atoms shared between the SiO$_4$ tetrahedra. Silicates with an olivine type stoichiometry  share no oxygen atoms and the bonds are formed through the cations. For magnesium/iron olivines this results in a chemical composition Mg$_{2x}$Fe$_{2-2x}$SiO$_4$, where $0<x<1$  determines the magnesium over iron ratio, $x=\mathrm{Mg/(Fe+Mg)}$. Silicates with a  pyroxene type stoichiometry share one oxygen atom, resulting in linked chains of  tetrahedra, which are bound by cations, resulting in a composition  Mg$_{x}$Fe$_{1-x}$SiO$_3$. When all oxygen atoms are shared between the tetrahedra we get  silica, SiO$_2$. Besides magnesium/iron silicates we also consider a sodium/aluminum  silicate with a pyroxene stoichiometry and equal amounts of Na and Al, i.e.  Na$_{0.5}$Al$_{0.5}$SiO$_3$. Since amorphous silicates have a disordered lattice  structure, we often find silicates in nature with a composition between the above  mentioned types. Mixing the silicates mentioned above results in an average composition  Mg$_{x_1}$Fe$_{x_2}$Na$_{x_3}$Al$_{x_3}$SiO$_{x_4}$, where $x_4=x_1+x_2+2x_3+2$. We will  use $x_4=\mathrm{O/Si}$ as a measure for the average stoichiometry. For olivines $x_4=4$,  for pyroxenes $x_4=3$ and for silica $x_4=2$.

In general the $10\,\mu$m feature for amorphous silicate with an olivine type  stoichiometry is broader and peaks at longer wavelengths than that of amorphous silicates  with a pyroxene type stoichiometry. For magnesium/iron silicates with much magnesium, i.e.  high values of $x$, the $10\,\mu$m feature is shifted to shorter wavelengths compared to  iron rich silicates, i.e. small values of $x$. The effect of composition on the shape of the 10\,$\mu$m silicate feature is illustrated  in Fig.~\ref{fig:Composition} for porous GRF particles.

In our analysis we also include crystalline silicates. As we mentioned earlier, the  crystalline silicates found in various astronomical environments are all magnesium rich,  i.e. $x\approx1$. We include the two most commonly found crystals namely crystalline  forterite (Mg$_2$SiO$_4$) and crystalline enstatite (MgSiO$_3$) in our analysis. Small  crystalline silicate grains display very strong spectral resonances that are easily  detectable in emission or extinction spectra.

Silicon carbide (SiC) is a dust material produced in carbon rich AGB stars \citep[see  e.g.][]{1997MNRAS.288..431S}. Furthermore, presolar SiC grains have been found in  meteorites and IDPs \citep[see e.g.][]{1987Natur.330..728B}. SiC displays a strong  spectral resonance around $11\,\mu$m. Therefore, we include the possibility of SiC in our  analysis.

An overview of all dust materials included in our analysis with references to the  laboratory data used to compute the extinction spectra can be found in  Table~\ref{tab:Materials}.

\begin{table}[!t]
\begin{center}
\begin{tabular}{lccc}
\hline
Name                & Composition   & Lattice Structure     & Ref.\\
\hline
Olivine ($x=0.5$)   & MgFeSiO$_4$       & Amorphous         & [1]\\
Pyroxene ($x=0.5$)  & MgFeSi$_2$O$_6$   & Amorphous         & [1]\\
Olivine ($x=1$)     & Mg$_2$SiO$_4$     & Amorphous         & [2]\\
Pyroxene ($x=1$)    & MgSiO$_3$         & Amorphous         & [1]\\
Na/Al Pyroxene      & NaAlSi$_2$O$_6$   & Amorphous         & [3]\\
Silica              & SiO$_2$           & Amorphous         & [4]\\
Forsterite          & Mg$_2$SiO$_4$     & Crystalline       & [5]\\
Enstatite           & MgSiO$_3$         & Crystalline       & [6]\\
Silicon Carbide     & SiC               & Crystalline       & [7]\\
\hline
\end{tabular}
\end{center}
\caption{The materials used in the fitting procedure. In the case of the amorphous silicates we use the names 'olivine' or 'pyroxene' to indicate the average silicate stoichiometry. For the coating of the coated GRF particles we use amorphous carbon as measured by \citet{1993A&A...279..577P}. The references in the last column 
refer to:
[1] \citet{1995A&A...300..503D},
[2] \citet{1996A&A...311..291H}
[3] \citet{1998A&A...333..188M},
[4] \citet{1960PhRv..121.1324S},
[5] \citet{Servoin},
[6] \citet{1998A&A...339..904J}, and
[7] \citet{1993ApJ...402..441L}.}
\label{tab:Materials}
\end{table}

\section{Interstellar silicate grains}
\label{sec:ISM grains}

\subsection{Observations}

\begin{figure}[!t]
\resizebox{\hsize}{!}{\includegraphics{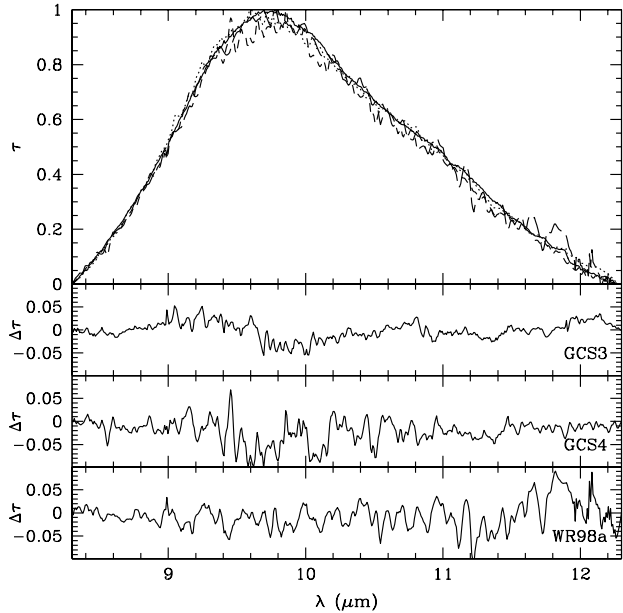}}
\caption{The normalized extinction profile in various lines of sight. A straight continuum  is subtracted such that at 8.3 and 12.3$\,\mu$m the extinction is 0. The features are  normalized such that the maximum extinction equals unity. In the upper panel we overlay  the extinction features towards Sgr A$^*$, GCS3, GCS4 and WR98a. In the three panels below  this we show the latter three extinction profiles minus the profile  observed towards Sgr A$^*$.}
\label{fig:Lines}
\end{figure}

The optical depth as a function of wavelength for interstellar dust grains can be  determined by observing a source located behind significant amounts of interstellar dust,  for example near the galactic center. When the spectrum of the observed source is known  the optical depth at each wavelength point, $\tau_\lambda$, of the dust grains can, in principle, be determined. In  practice, the spectrum of the source is often not known, and it is only possible to derive the optical depth in the silicate feature itself. This was done for the  line of sight towards Sgr A$^*$ by \citet{2004ApJ...609..826K}, and we will use their  results in our analysis. Although we restrict ourselves to this line of sight, the  extinction features in other lines of sight towards the galactic center look very similar  \citep[see Fig.~\ref{fig:Lines}; cf.][]{2004ApJ...609..826K}. We use the feature  towards Sgr A$^*$ because it has the best signal to noise.

\subsection{Fitting procedure and error analysis}

We fit the interstellar 10$\,\mu$m extinction feature by making a linear least square fit to the optical depth in the feature using the extinction spectra computed for the various dust components  listed in Table~\ref{tab:Materials}. 
In order to correct for the continuum opacity in the computed extinction spectra of the dust species we subtract a straight line from the fit. We do this by introducing two additional fit parameters to the fitting procedure, the slope and offset of this line. In this way we treat the observations and the opacities in a consistent manner. We would like to stress here that by this procedure we fit the optical depth in the feature, and not the total optical depth caused by the dust species.
In order to get an idea of the errors we make by assuming a straight line continuum, we have also made fits using different power law continua. This only had a very small influence on the results and did not affect any of our conclusions.


To avoid fits with an unlikely dust composition due to unrealistic combinations of elemental abundances we fix  the elemental abundance of Mg relative to Si. Similar constraints were used by  \citet{2004ApJS..152..211Z} to fit the interstellar extinction profile. To do this we need  a good estimate for the abundances of Mg and Si in the ISM. \citet{2004ApJS..152..211Z}  use the average abundances of F-, G- or B-type stars to constrain their model. We use the  $\mathrm{Mg/Si}$ ratio as derived by \citet{2005ApJ...620..274U} from X-ray spectroscopy.  Using their technique they probe simultaneously the elements in the gas phase and the  solid phase of the ISM. They derive $\mathrm{Mg/Si}=1.22\pm0.14$, where the error reflects a 90\% confidence limit (which corresponds to $\sim1.65\sigma$). We use a $1\sigma$ error implying $\mathrm{Mg/Si}=1.22\pm0.09$. We assume all Mg  and Si to be in the solid phase consistent with the results of \citet{2005ApJ...620..274U}.

In order to get an idea of the error on the parameters fitted to the interstellar  10$\,\mu$m extinction feature we generated 1000 different synthetic observations. For the  extinction spectrum we did this by adding noise to the spectrum according to the error on  the spectrum at each wavelength point, i.e. $\sigma_{\lambda}$. The error on the spectrum as  derived by \citet{2004ApJ...609..826K} is very small. The large value of $\chi^2$ obtained from the spectral fit by \citet{2004ApJ...609..826K} already indicates this.  This is probably due to deficiencies in the model. To account for this underestimate of the error we scaled  the $\sigma_{\lambda}$ by a factor $\sqrt{\chi^2}$ and used this scaled error to construct  the 1000 synthetic spectra to be fitted. In this way when the fit to the spectrum is poor,  i.e. a high value of $\chi^2$ is obtained, the errors on the derived fit parameters will also be  large. This reflects that a high value of $\chi^2$ indicates that the model does not  accurately describe the observations.

For each spectrum to be fitted we also constructed a synthetic abundance constraint using  $\mathrm{Mg/Si}=1.22\pm0.09$. So we employ a slightly different abundance constraint for each  synthetic spectrum. For each of the 1000 combinations of spectra and abundance constraints  we applied our fitting procedure to obtain the abundances of the various dust species.  This results in a distribution of abundances for all components. This distribution can be used  to compute the average fit parameters along with an estimate of the standard deviations on  these parameters.

The errors obtained by the above described procedure are the errors on the abundances {\em assuming the dust component is present.} A better measure for the significance of the detection of a dust component can be obtained by performing a so-called F-test. This test determines the significance of a component in the fit by measuring the decrease of the $\chi^2$ when the component is added. A simple F-test, as usually employed, is only valid under strict assumptions of the distribution of $\chi^2$ under variation of the observations. However, fitting the 1000 synthetic spectra allows us to create the F-distribution exactly \citep[see][]{Chernick}. For each component we also make a fit to all the spectra leaving out this component. This gives two distributions of $\chi^2$, one with and one without the specific component. The distribution of the ratios of elements from these two distributions is called the F-distribution. By determining the mean and the standard deviation, $\sigma$, of this distribution, we can determine how far (in terms of a multiple of $\sigma$) the mean lies from unity. Since a mean value of unity implies that the fit is insensitive for this specific component, the deviation from unity gives the significance of the detection. For more details we refer to \citet{Chernick}. In order to minimize the number of parameters in the fit and in this way avoid over interpretation of the observations, we iteratively remove the least significant component until all components are detected with at least $1\sigma$.

To summarize the fitting procedure:
\begin{enumerate}
\item Make a linear least squares fit to the observed $\tau_\mathrm{feature}$ using the  abundance constraint and the errors on the spectrum, $\sigma_{\lambda}$, as determined by  \citet{2004ApJ...609..826K}. This yields the reported $\chi^2$.
\item Construct $\sigma'_{\lambda}$ so that the best fit using these errors would have a  $\chi^2$ of unity, i.e. $\sigma'_{\lambda}=\sigma_{\lambda}\,\sqrt{\chi^2}$.
\item Use these $\sigma'_{\lambda}$ and the errors on the abundance constraint to  construct 1000 combinations of synthetic spectra and abundances by adding Gaussian noise  to the observations.
\item Make linear least squares fits to all these synthetic data sets.
\item For all components in the fitting procedure make linear least squares fits to all synthetic data sets leaving out this component.
\item Use the fit parameters of all 1000 fits to obtain the average abundances and use the $\chi^2$ distributions obtained in steps 4 \& 5 to construct the F-distributions. The F-distributions are used to derive the significances of all dust components.
\item If there are dust components detected with a significance less than $1\sigma$, remove the least significant component and restart the fitting procedure from point 1. Iterate until all components left in the procedure are detected with at least $1\sigma$.
\end{enumerate}

\begin{figure*}[!t]
\resizebox{\hsize}{!}{\includegraphics{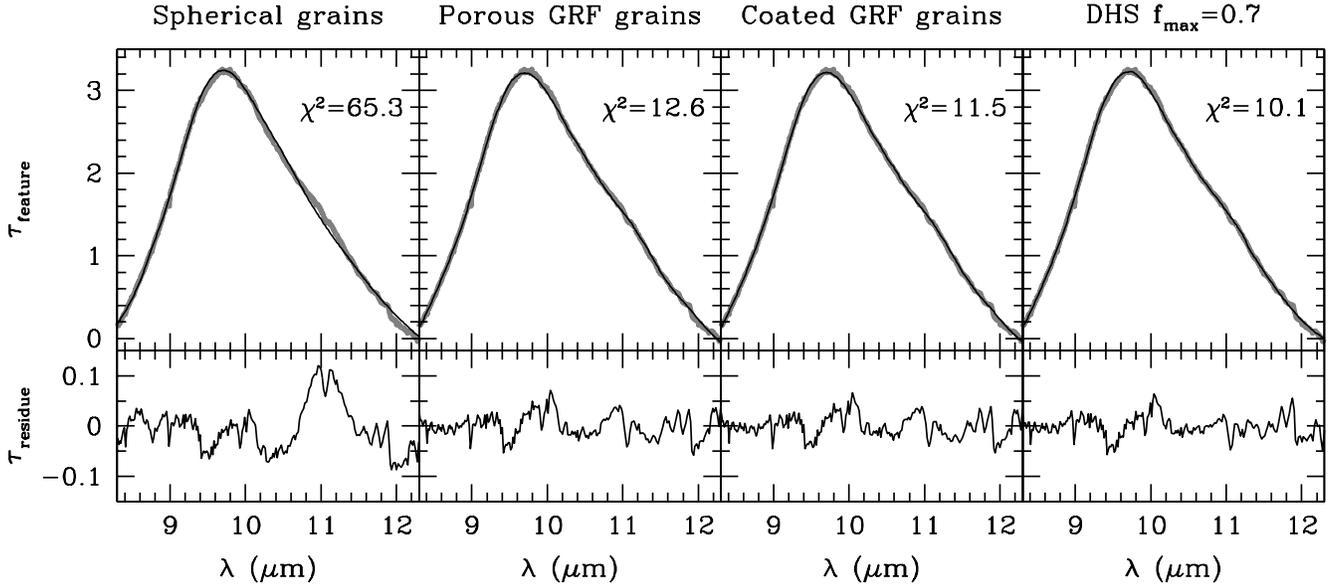}}
\caption{The wavelength dependence of the extinction of the best fit models (black curves) together with the  extinction feature as obtained by \citet{2004ApJ...609..826K} (gray curves). Here, $\tau_\mathrm{feature}$ denotes the optical depth in the 10$\,\mu$m feature, and $\tau_\mathrm{residue}$ denotes the observed optical depth minus the modeled optical depth.}
\label{fig:ISM Fit}
\end{figure*}

\begin{table*}[!t]
\begin{center}
\linespread{1.3}%
\selectfont
\begin{tabular}{lcrcrcrcrc}
\hline
                &             & \multicolumn{2}{c}{\parbox{2.5cm}{\centerline{Spherical grains}}} & \multicolumn{2}{c}{\parbox{2.5cm}{\centerline{Porous GRF grains}}}  & \multicolumn{2}{c}{\parbox{2.5cm}{\centerline{Coated GRF grains}}}  & \multicolumn{2}{c}{\parbox{2.5cm}{\centerline{DHS ($f_\mathrm{max}=0.7$)}}}   \\
\multicolumn{2}{c}{$\chi^2$}  & \multicolumn{2}{c}{65.3}          & \multicolumn{2}{c}{12.6}              & \multicolumn{2}{c}{11.5}              & \multicolumn{2}{c}{10.1}                       \\
\hline\hline
\multicolumn{2}{l}{Silicates}\\
\hline
Olivine (A; $x=0.5$)   & MgFeSiO$_4$       &\qquad\quad 57.9   & (6.2$\sigma$)   &\qquad\quad16.9   & (1.9$\sigma$)    &\qquad\quad 9.7   & (1.1$\sigma$)  &\qquad\quad 13.8   & (2.0$\sigma$)   \\
Pyroxene (A; $x=0.5$)  & MgFeSi$_2$O$_6$	& 12.9	& (7.6$\sigma$)	&  -		&				&  -		&				&  -		&    \\
Olivine (A; $x=1$)     & Mg$_2$SiO$_4$		& 28.9	& (1.0$\sigma$)	& 47.2	& (3.5$\sigma$)	& 48.2	& (3.5$\sigma$)	& 38.3	& (3.8$\sigma$)    \\
Pyroxene (A; $x=1$)    & MgSiO$_3$		&  -		&				& 24.0	& (6.1$\sigma$)	& 31.4	& (6.8$\sigma$)	& 42.9	& (6.7$\sigma$)    \\
Na/Al Pyroxene (A)     & NaAlSi$_2$O$_6$	&  -		&    				&  6.8  	& (3.2$\sigma$)   	&  5.3  	& (2.7$\sigma$)   	&  1.8  	& (1.1$\sigma$)    \\
Silica (A)             & SiO$_2$           			&  0.3  	& (1.1$\sigma$)	&  -  		&   	 			&  -  		&    				&  -  		&   \\
Forsterite (C)         & Mg$_2$SiO$_4$     		&  -  		&    				&  1.5  	& (4.7$\sigma$)   	&  1.3  	& (3.6$\sigma$)   	&  0.6  	& (3.2$\sigma$)    \\
\hline
\multicolumn{2}{l}{Other}\\
\hline
Silicon Carbide (C)    & SiC           			&  -  		&    				&  3.6  	& (3.7$\sigma$)   	&  4.2  	& (4.3$\sigma$)   	&  2.6  	& (5.3$\sigma$)    \\
\hline
\multicolumn{8}{l}{The average silicate composition implied by the above abundances}\\
\hline
$x_1$       & Mg/Si		& 1.22  &   & 1.37  &   & 1.39  &   & 1.32  &   \\
$x_2$       & Fe/Si		& 0.60  &   & 0.13  &   & 0.07  &   & 0.10  &   \\
$x_3$       & (Na, Al)/Si	& 0.00  &   & 0.05  &   & 0.03  &   & 0.01  &   \\
$x_4$       & O/Si		& 3.82  &   & 3.59  &   & 3.53  &   & 3.45  &   \\
\hline
\end{tabular}
\linespread{1}%
\selectfont
\end{center}
\caption{The composition in terms of mass fractions in percent together with the significance of the detection of the dust components for the best fit models using various 
particle shapes. A '-' denotes that the dust component was detected with a significance less than 1$\sigma$ and was thus removed from the fitting procedure. Enstatite was not found to be significant using any of the four grain shapes and is therefore omitted from the table. The (A, C) behind the name in the first column denotes if the material is amorphous or crystalline. The names 'olivine' and 'pyroxene' are used for the amorphous silicates to refer to the average stoichiometry. The $x_1, x_2, x_3, x_4$ refer to the average silicate composition, Mg$_{x_1}$Fe$_{x_2}$Na$_{x_3}$Al$_{x_3}$SiO$_{x_4}$ implied by the abundances of all silicates.}
\label{tab:Composition}
\end{table*}

\subsection{Results from the 10 micron feature}

The best fits to the 10$\,\mu$m feature using homogeneous spheres, porous GRF particles,  coated porous GRF particles, and a Distribution of Hollow Spheres are shown in Fig.~\ref{fig:ISM Fit}. For the DHS we determined that the  best fit was obtained by taking $f_\mathrm{max}=0.7$, similar to the best fit value found  in the analysis of comet Hale-Bopp \citep{2005Icar..179..158M}. The derived composition  for each of these shape distributions is listed in Table~\ref{tab:Composition}. The  $\chi^2$ of these fits show that the feature is better represented by using irregularly shaped  particles than by using homogeneous spherical particles.

\begin{figure}[!t]
\resizebox{\hsize}{!}{\includegraphics{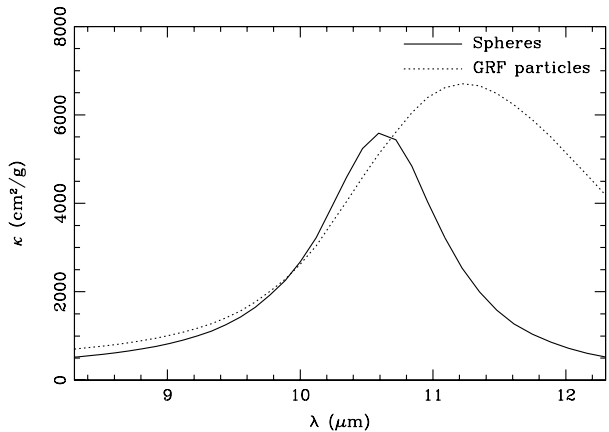}}
\caption{The mass extinction coefficient as a function of wavelength in the $10\,\mu$m  region for silicon carbide (SiC) particles with various shape distributions.}
\label{fig:SiC spectra}
\end{figure}

The resulting dust composition using irregularly shaped particles is very different from  that obtained when using homogeneous spheres. 
Since it seems unlikely that the interstellar silicates are perfect homogeneous spheres, and there are no other particle shapes that produce a spectrum similar to that of perfect homogeneous spheres, this model must be rejected. In addition, the fit using homogeneous spheres has a reduced $\chi^2$ which is almost six times as high as those obtained using other particle shapes.
From the analysis we draw the following conclusions.

i) The interstellar silicate grains contain much more magnesium than previously assumed. From  Table~\ref{tab:Composition} it can be deduced that when using the porous GRF particles or  the DHS, we derive that the magnesium content $\mathrm{Mg/(Fe+Mg)}=0.91$ and $0.93$ respectively. Using homogeneous  spheres we derive $\mathrm{Mg/(Fe+Mg)}=0.67$. The fit using coated GRF particles even results in $\mathrm{Mg/(Fe+Mg)}=0.95$.

ii) We confirm the findings by \citet{2004ApJ...609..826K, 2005ApJ...633..534K} that the  fraction of crystalline silicates (i.e. crystalline forsterite and enstatite) in the ISM  is very low ($\sim0.6-1.5$\%).

iii) We find a significant fraction ($\sim$2.6-4.2\%) of silicon carbide (SiC). This implies that $\sim$9-12\% of the available Si is in these SiC grains. \citet{2004ApJ...609..826K} placed an upper limit of 0.1\% on the abundance of homogeneous  spherical SiC grains in the ISM from their analysis of the 10$\,\mu$m spectrum. However, the position and shape of the SiC resonance in the 10$\,\mu$m region is very sensitive to the particle shape \citep[see e.g.][]{2005astro.ph.11371A} and size \citep[see][]{2005ApJ...634..426S}. Whereas  homogeneous spherical SiC grains show a sharp resonance at 10.6$\,\mu$m, irregularly  shaped SiC particles show a much broader spectral feature peaking near 11.25$\,\mu$m \citep[see  Fig.~\ref{fig:SiC spectra} or][]{2005ApJ...634..426S}. The observed 10$\,\mu$m extinction spectrum  (Fig.~\ref{fig:ISM Fit}) has a shoulder around 11$\,\mu$m which is fitted  correctly using(coated) porous GRF or DHS particles but which is not fitted at all using  homogeneous spheres. This is due to the presence of irregularly shaped SiC grains. 
The  presence of SiC in the ISM is not unexpected since presolar SiC grains have been found in  dust grains in the Solar system \citep{1987Natur.330..728B}. These SiC grains found in  IDPs are indeed irregularly shaped. Implications of the presence of SiC in the diffuse ISM will be discussed in a separate study.

iv) We find that the average stoichiometry of the silicates corresponds to  $\mathrm{Si/O}\approx3.5$ when applying irregularly shaped particles, i.e. between an  olivine and a pyroxene type stoichiometry. When using homogeneous spherical particles we  find a stoichiometry more weighted towards olivine type silicates  \citep[cf.][]{2004ApJ...609..826K}. The implications for these findings will be discussed  in section~\ref{sec:Discussion}.

\subsubsection{The SiC lattice structure}
\label{sec:SiC discuss}

\begin{table*}[!t]
\begin{center}
\linespread{1.3}%
\selectfont
\begin{tabular}{lcccccccc}
\hline
		& \multicolumn{2}{c}{Spherical grains} & \multicolumn{2}{c}{Porous GRF grains} &  \multicolumn{2}{c}{Coated GRF grains} & \multicolumn{2}{c}{DHS ($f_\mathrm{max}=0.7$)}\\
\hline
Crystalline SiC (Laor \& Draine 1993) 		&\multicolumn{2}{c}{-}&3.6\%	& (3.7$\sigma$)	&4.2\%	& (4.3$\sigma$)	&2.6\%	& (5.3$\sigma$)\\
Crystalline $\beta$-SiC (Mutschke et al. 1999)	&\multicolumn{2}{c}{-}&1.0\%	& (2.6$\sigma$)	&1.2\%	& (2.4$\sigma$)	&0.3\%	& (4.0$\sigma$)\\
Amorphous SiC (Mutschke et al. 1999)		&\multicolumn{2}{c}{-}&5.8\%	& (3.1$\sigma$)	&5.0\%	& (2.9$\sigma$)	&4.1\%	& (2.8$\sigma$)\\
\hline
\end{tabular}
\linespread{1}%
\selectfont
\end{center}
\caption{The results of the analysis using various SiC measurements available in the literature. We conclude that the SiC in the ISM has a lattice structure closest resembled by that of \citet{1993ApJ...402..441L}.
}
\label{tab:SiC}
\end{table*}

Silicon carbide can exist in many different crystal structures. \citet{1999A&A...345..187M} have studied the dependence of the spectral appearance of the 11$\,\mu$m resonance on crystal structure. They concluded that when the crystal is pure, the actual crystal structure is not very important. However, they did note a large difference when the crystal structure is destroyed and an amorphous SiC structure is left. To test the influence of the optical data taken to model the SiC resonance we considered three different sets of refractive indices. First we considered the data as shown before \citep{1993ApJ...402..441L}. Second we take a rather pure crystal ($\beta$-SiC with $\gamma=10\,$cm$^{-1}$) according to the equations given by \citet{1999A&A...345..187M}. Third we consider amorphous SiC as measured by \citet{1999A&A...345..187M}. The results of the different fits are shown in Table~\ref{tab:SiC}. From this analysis it seems that the most likely appearance of SiC in the ISM is closest to the measurements by \citet{1993ApJ...402..441L}. The abundances of the other dust components are not significantly different from those presented in Table~\ref{tab:Composition}.

\subsection{Comparison with the 20 micron silicate feature}

Apart from the frequently used 10$\,\mu$m feature, silicates also display a spectral signature that peaks near 20$\,\mu$m due to the O-Si-O bending mode. It is often referred to as the 18$\,\mu$m feature, although its spectral position varies with silicate composition and particle shape. We will refer to it as the 20$\,\mu$m feature. The spectral shape of this feature is very sensitive to particle shape and composition. Because it is rather broad, it is very hard to extract the 20$\,\mu$m extinction profile  accurately from measured infrared spectra. Only if the spectrum of the emission source  behind the column of extincting dust is known rather accurately, a reliable extraction of  the extinction profile can be made. This has been done by \cite{2006ApJ...637..774C} from  the spectrum taken by the \emph{Infrared Space Observatory} (ISO) towards WR~98a. This resulted in an extinction profile covering both the 10 and 20$\,\mu$m  silicate feature. The spectral signature derived by \cite{2006ApJ...637..774C} is much noisier than the 10$\,\mu$m feature we used for our analysis above. Therefore, we chose to do the detailed analysis on the 10$\,\mu$m feature and then check if the resulting dust composition and shape distribution is consistent with the full spectral profile.

\begin{figure}[!t]
\resizebox{\hsize}{!}{\includegraphics{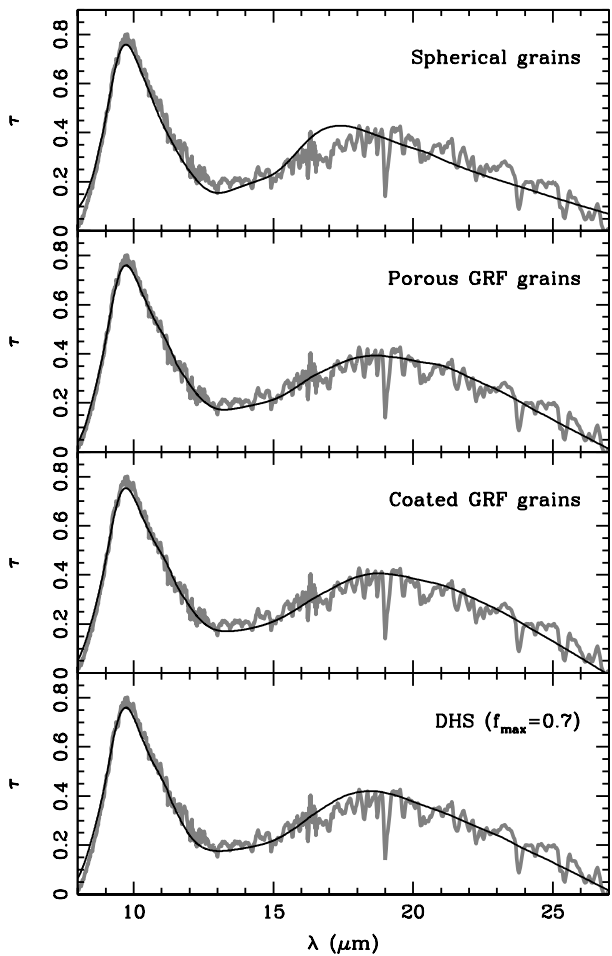}}
\caption{The extinction spectra of the best fit models computed over a longer wavelength  range (black curves) and compared to the feature obtained by \citet{2006ApJ...637..774C}  (gray curves).}
\label{fig:20 micron}
\end{figure}

In order to check whether the composition and shape distribution obtained from the fit to  the 10$\,\mu$m feature is consistent with the observed 20$\,\mu$m feature, we computed the  expected extinction profile over the total spectral range for the dust composition and  shape distribution as derived from the 10$\,\mu$m feature. To account for possible errors  in the intrinsic spectrum of the source used to extract the dust extinction profile from the observations, we added a smooth  quadratic continuum to the computed extinction profile in order to fit the observed  extinction spectrum. We do not think that the exact choice of the continuum has a significant impact on our results. The results are shown in Fig.~\ref{fig:20 micron}. It is clear from  this figure that the spectra using (coated) porous GRF or DHS particles give a much better fit to  the observed spectral profile than the spectrum using homogeneous spheres. This is  consistent with the conclusion by \cite{2006ApJ...637..774C} that perfect homogeneous  spheres cannot account for the observed 20$\,\mu$m feature.

\section{Implications}
\label{sec:Discussion}

\subsection{Processing of dust in protoplanetary disks}

An important unsolved issue is that of the apparent difference in the composition of crystalline and amorphous silicates in protoplanetary disk material: silicate crystals are found to be magnesium rich and iron poor, while amorphous silicates have been anticipated to have more or less equal amounts of iron and magnesium. As it is usually assumed that the bulk of the crystalline silicates in disks are formed by thermal annealing of amorphous silicates, it is difficult to reconcile their different chemical composition. The composition of the crystalline component is a firm result since emission spectra from crystalline  silicates show very narrow emission features, the wavelength of which is a strong  indicator of the magnesium content. Cometary and  circumstellar crystalline silicates are almost pure magnesium silicates \citep[see  e.g.][]{1998A&A...339..904J, 2001A&A...378..228F}. Such magnesium silicate crystals are  expected to form from direct gas phase condensation \citep{1972GeCoA..36..597G}. Using  radial mixing and thermal processing of grains \citet{2004A&A...413..571G} provides  predictions of the composition and lattice structure of silicate grains in the region  where in the solar system the comets have assembled. From these predictions it follows  that the crystalline silicates in this region should be only partly magnesium rich -- the  part which is formed from direct gas phase condensation very close to the star -- and  partly still reflect the composition of the original amorphous silicates -- the part which  is formed by thermal annealing of amorphous silicates. The magnesium rich gas phase  condensates are only expected to make up $\sim10$\% of the crystalline silicates in the  regions where the comets have assembled and thus cannot explain the high magnesium  fraction in cometary crystals.
We have shown, however, that the interstellar amorphous silicates are already very rich in  magnesium. Thus the amorphous material from which the crystalline silicates form is  already magnesium rich. This provides a very natural explanation for the composition of  crystalline silicates in cometary and protoplanetary dust. Indeed, \citet{2006A&A...448L...1D} show that when an amorphous silicate with $\mathrm{Mg/(Fe+Mg)}=0.9$ is annealed, the resulting crystalline silicates are almost purely magnesium rich.

\subsection{Depletion of elements along various lines of sight}

If the interstellar silicates are magnesium rich, the question arises: where is the iron? Unusual depletion of Si, Fe and Mg of lines of sight in the Small  Magellanic Cloud have been reported \citep{2001ApJ...554L..75W, 2006ApJ...636..753S}. In these lines of  sight most Si and Mg are in the gas phase indicating that the silicate component of the  interstellar dust is destroyed. However, the iron depletion is unchanged, i.e. most of the  iron is still in the solid phase. This indicates that iron and magnesium are in different  dust materials as suggested already by \citet{2006ApJ...636..753S}. If the iron would be incorporated in the silicate lattice it would return  to the gas phase along with the magnesium and silicon when the silicate is destroyed.  Our analysis indeed shows that the silicate is highly magnesium rich indicating that the iron must be in the form of refractory species such as metallic iron or iron oxide. In this scenario, when the silicate is  destroyed, the iron can stay in the solid phase while the magnesium and silicon are  returned to the gas phase. This thus provides a natural explanation of the unusual depletion  patterns reported.

\subsection{The nature of GEMS}

It has been suggested that GEMS (Glass with Embedded Metal and Sulphides), which are small  subgrains found in many IDPs, are unprocessed interstellar dust grains  \citep{1999Sci...285.1716B}. This raises the question: is the chemical composition of ISM grains that we have derived consistent with that of GEMS? The idea that GEMS are unprocessed interstellar grains is supported by a similarity between the shape of  the 10$\,\mu$m feature of some GEMS and that observed for the interstellar medium \citep{1999Sci...285.1716B}, although also distinct differences can be identified \citep{1998LPI....29.1737B}. In addition, traces of exposure of GEMS to ionizing radiation consistent with an interstellar  origin have been reported \citep{1994Sci...265..925B}. Also, for some individual GEMS it is determined that they are of extrasolar origin \citep[see e.g.][]{2003Sci...300..105M, 2006GeCoA..70.2371F}.

The amorphous  (glassy) silicate in GEMS has a magnesium content of $\mathrm{Mg/(Fe+Mg)}\approx 0.9$, consistent with our analysis of ISM grains. The iron found in GEMS is mainly in the form of inclusions of FeS and metallic Fe, only a minor fraction is inside the silicate lattice. Among these components iron sulphide (FeS) is  dominant. This is in contrast with studies of depletion of sulfur from the gas phase in the diffuse interstellar medium which show that the sulfur in the ISM is predominantly in the gas phase \citep[see e.g.][]{2005ApJ...620..274U,  1996ARA&A..34..279S, 1996ApJ...468L..65S}. Preliminary observations indicate that the iron  sulphide inclusions in GEMS are located preferably at the edge of the grains \citep{2005LPI....36.2088K}. This might indicate that metallic iron inclusions inside ISM grains reacted  with sulfur to form the FeS inclusions in GEMS. If so, this process must have happened in  the collapsing molecular cloud, or in the protoplanetary disk phase. 

The average GEMS silicate has  $\mathrm{(Mg+Fe)/Si}\approx 0.7$ \citep{2004LPI....35.1985K}, much lower than we derive  ($\mathrm{(Mg+Fe)/Si}\approx 1.5$, see Table~\ref{tab:Composition}). This gives $\mathrm{O/Si}\approx 2.7$ in the GEMS  silicates, implying they are, on average, almost pure pyroxene with a small amount of  silica. This difference is consistent with the spectral position of the maximum absorption reported for GEMS which is around 9.3$\,\mu$m \citep{1999LPI....30.1835B}, while it is around 9.7$\,\mu$m in the ISM. Considering the fact that decreasing the $\mathrm{O/Si}$ in the amorphous silicates tends to shift the maximum absorption towards shorter wavelengths (see Fig.~\ref{fig:Composition}), we can understand this difference between the GEMS and ISM 10$\,\mu$m features in terms of a difference in average stoichiometry. If we apply abundance constraints as determined for GEMS in our fitting of the 10$\,\mu$m feature we get no satisfying fit ($\chi^2>90$ for all particle shape models).  The arguments above taken together suggest that GEMS are, in general, not unprocessed leftovers from  the diffuse ISM. However, they may have formed in the collapsing molecular cloud from which the Solar  system was formed.

\section{Conclusions}
\label{sec:Conclusions}

We have shown that the interstellar silicate extinction spectrum can be fitted accurately  using irregularly shaped particles. In previous studies the silicate extinction has been explained  using spherical amorphous silicate grains with equal amounts of iron and magnesium.  However, interstellar silicates are most likely not perfect homogeneous  spherical grains. Using irregularly shaped dust grains we find a reduced $\chi^2$ that is almost six times smaller than that obtained using homogeneous spheres. Moreover, the spectrum of homogeneous spherical grains is inconsistent with the observed 20$\,\mu$m extinction feature. Using irregular particle shapes, we analyze the 10$\,\mu$m extinction feature in detail. The resulting dust composition and shape distribution is also found to be consistent with the observed 20$\,\mu$m feature.

To summarize we conclude the following concerning the composition of the interstellar  silicates:
\begin{itemize}
\item The amorphous silicates in the ISM are highly magnesium rich. We show that the average composition of the silicates corresponds to $\mathrm{Mg/(Fe+Mg)}>0.9$.
\item We confirm the findings of previous studies that the crystallinity of the silicates in the ISM is small, although we do find a small fraction of crystalline forsterite. We find that the crystallinity is approximately 1\%.
\item We find a mass fraction of approximately 3\% of silicon carbide. The presence of SiC in the ISM has been expected since presolar SiC grains are frequently found in IDPs. We identify the presence of SiC grains in the ISM spectroscopically with more than 3$\sigma$ confidence.
\item The average stoichiometry of the interstellar silicates is between olivine and pyroxene type silicates. We find that $\mathrm{O/Si}\approx3.5$.
\end{itemize}

Our analysis has several implications for the interpretation of silicate dust in various  environments. These are:
\begin{itemize}
\item The high magnesium content of amorphous silicates in the ISM provides a natural explanation for the high magnesium content of crystalline silicates found in cometary and circumstellar grains. We argue that these magnesium rich crystals can be formed by simple thermal annealing of the magnesium rich amorphous silicates.
\item Unusual depletion patterns observed in several lines of sight through the interstellar medium show that the depletion of Fe is not correlated with that of Mg and Si. This is naturally explained when these components are not in the same dust materials, consistent with our findings.
\item We discuss the origin of GEMS in interplanetary dust particles. These grains are suggested to be of interstellar origin. However, we show that the composition of the ISM silicates is not consistent with that of GEMS, suggesting that GEMS are not unprocessed remnants from the ISM.
\end{itemize}

As a concluding remark we would like to point out that the use of homogeneous spherical particles to model silicate extinction  spectra gives results that deviate in important aspects from the values obtained using more realistic  particle shapes. Since most dust grains in nature are not homogeneous spheres the use of homogeneous spherical particles should be avoided, even if they do  provide a reasonable fit to the observations as it may yield spurious grain properties. In cases where modeling of realistically  shaped particles is computationally too time consuming, the Distribution of Hollow Spheres  provides a good alternative. We have shown that this shape distribution simulates the  properties of irregularly shaped particles, while computational effort is small.

\begin{acknowledgements}
We are grateful to J.~E. Chiar for providing us with the full extinction profile of the  interstellar silicates. M.  Min acknowledges financial support from the Netherlands Organisation for Scientific Research (NWO) through a Veni grant. We would like to thank an anonymous referee for constructive comments.
\end{acknowledgements}

\appendix
\section{Construction of porous Gaussian Random Field particles}
\label{app:GRF}

In order to construct the Gaussian Random Field particles (hereafter GRF particles), we employed the following algorithm. First we constructed a three-dimensional field with random numbers $R_{ijk},\,(i,j,k=1..M)$. From this we constructed a field $G_{ijk}$ according to
\begin{equation}
G_{ijk}=R_{ijk}\,e^{-\rho\,d^2}
\end{equation}
where $\rho$ determines the size of the particles to be formed and
\begin{equation}
d=\sqrt{(i-M/2)^2+(j-M/2)^2+(k-M/2)^2}
\end{equation}
is the distance to the center of the field. The normalized Gaussian Random Field is obtained from $G_{ijk}$ by taking its three-dimensional Fourier transform
\begin{equation}
\mathcal{G}_{ijk}=\mathcal{F}\left(G_{ijk}\right)/\max_{ijk}\mathcal{F}\left(G_{ijk}\right).
\end{equation}
where $\mathcal{F}$ denotes the Fourier transform. From this normalized three-dimensional field $\mathcal{G}_{ijk}$ we construct the particles by taking all points in space where $\mathcal{G}_{ijk}>0.5$ to be inside the particle, while $\mathcal{G}_{ijk}<0.5$ is taken to be outside the particle. In this way $\mathcal{G}_{ijk}$ represents a field filled with a large number of particles. We have to separate these in order to get single particles. 

For the computation of the extinction spectra of these particles we use the Discrete Dipole Approximation (DDA). In the DDA we approximate the particle by discretizing the particle volume into small volume elements that each interact with the incoming light as dipoles. The $\mathcal{G}_{ijk}$ immediately provides a grid for these dipoles. For the final DDA calculations we only selected those particles that contain between 2000 and 3500 dipoles.

In order to construct porous Gaussian Random Field particles, we make vacuum inclusions in the particles by computing a second Gaussian Random Field, 
$\mathcal{G}'_{ijk}$, with a much smaller effective size, $\rho'$. The points in space where this second field has values larger than a certain threshold are considered vacuum. By changing the threshold value and the size of the vacuum inclusions, $\rho'$, we can vary the degree of porosity and irregularity of the final particles. We have taken $\rho'=\rho/50$ and a threshold value of $0.1$.

The coated porous GRF particles are created in the same manner as the pure GRF particles but we take another threshold value below which the material is considered mantle material. In this way the outer regions of the grain become the mantle, while the inner regions of the grain (where the GRF has its highest values) consists of the core material. This threshold is chosen such that each particle is 'covered' with a mantle containing 30\% of the material volume of the grain.

We constructed an ensemble of $38$ porous GRF particles and averaged the optical properties over this ensemble. Typical examples of the particles are shown in Fig.~\ref{fig:GRF particles}. An example of a coated porous GRF particle is shown in Fig.~\ref{fig:CoatedGRF}.


\end{document}